\documentclass[manuscript]{aastex61}

\newcommand\aastex{AAS\TeX}

\shorttitle{\aastex\ artical}
\shortauthors{Wang et al.}

\begin{document}

\title{Roles of photospheric motions and flux emergence in the major solar eruption on 2017 September 6}

\author{Rui Wang}
\affil{State Key Laboratory of Space Weather, National Space Science Center, Chinese Academy of Sciences, Beijing, China}
\affil{W.W. Hansen Experimental Physics Laboratory, Stanford University, Stanford, CA, USA}
\author{Ying D. Liu}
\affiliation{State Key Laboratory of Space Weather, National Space Science Center, Chinese Academy of Sciences, Beijing, China}
\affil{University of Chinese Academy of Sciences, Beijing 100049, China}

\author{J. Todd Hoeksema}
\affil{W.W. Hansen Experimental Physics Laboratory, Stanford University, Stanford, CA, USA}
\author{I.V. Zimovets}
\affil{State Key Laboratory of Space Weather, National Space Science Center, Chinese Academy of Sciences, Beijing, China}
\affil{International Space Science Institute -- Beijing (ISSI-BJ), Beijing 100190, China}
\affil{Space Research Institute (IKI) of the Russian Academy of Sciences, Moscow 117997, Russia}
\author{Yang Liu}
\affil{W.W. Hansen Experimental Physics Laboratory, Stanford University, Stanford, CA, USA}

\begin{abstract}
We study the magnetic field evolution in the active region (AR) 12673 that produced the largest solar flare in the past decade on 2017 September 6. Fast flux emergence is one of the most prominent features of this AR. We calculate the magnetic helicity from photospheric tangential flows that shear and braid field lines (shear-helicity), and from normal flows that advect twisted magnetic flux into the corona (emergence-helicity), respectively. Our results show that the emergence-helicity accumulated in the corona is $-1.6\times10^{43}~Mx^2$ before the major eruption, while the shear-helicity accumulated in the corona is $-6\times10^{43}~Mx^2$, which contributes about 79\% of the total helicity. The shear-helicity flux is dominant throughout the overall investigated emergence phase. Our results imply that the emerged fields initially contain relatively low helicity. Much more helicity is built up by shearing and converging flows acting on preexisted and emerging flux. Shearing motions are getting stronger with the flux emergence, and especially on both sides of the polarity inversion line of the core field region. The evolution of the vertical currents shows that most of the intense currents do not appear initially with the emergence of the flux, which implies that most of the emerging flux is probably not strongly current-carrying. The helical magnetic fields (flux rope) in the core field region are probably formed by long-term photospheric motions. The shearing and converging motions are continuously generated driven by the flux emergence. AR 12673 is a representative as photospheric motions contribute most of the nonpotentiality in the AR with vigorous flux emergence.

\end{abstract}


\section{INTRODUCTION}
Magnetic fields are believed to play a fundamental role in producing solar flares and coronal mass ejections (CMEs) that have drawn the majority of scientific interest due to their significant impact on the space weather. Magnetic fields form magnetic structures in the solar coronal which are associated with plasma motions in the photosphere. Photospheric motions can perturb the magnetic structures by shearing or converging field lines relatively to the polarity inversion line (PIL; e.g., \citealp{2001Moore,1989Vanballegooijen}), thereby drive the magnetic reconnection in the corona. Photospheric shearing motions are one of the possible way to introduce strong electric currents and inject energy into the active region (AR) corona (e.g, \citealt{2012SunXD}). Photospheric motions are generally formed due to convection and differential rotation. In an emerging flux region, horizontal divergent flows are often observed before the flux emergence. \citet{2012Toriumi} found the horizontal flows are associated with the rising flux tube below the photosphere. They indicated that as the rising tube approaches the photosphere, the accumulated plasma on the rising tube escapes horizontally around the surface and is observed as a horizontal divergent flow. On the other hand, simulations show that emergence of sufficiently twisted flux rope structures can also drive photospheric motions (e.g., rotational or shearing motions; \citealt{2004Manchester,2006Magara,2007Manchester,2009FanYH,2012Fangfang}). The basic underlying reason is the nature of the Lorentz force. \citet{2003Magara} by magnetohydrodynamics (MHD) numerical simulation found that flux emergence generates not only vertical but also horizontal flows in the photosphere, both of which contribute to injecting magnetic energy and magnetic helicity. The contributions of vertical flows are dominant at the early phase of flux emergence, while horizontal flows become a dominant contributor later. Consistently, in the early emergence phase, the flux of helicity and magnetic energy are both dominated from the contributions of vertical flows that advect twisted magnetic flux into the corona; in the later phases of flux emergence, both quantities are dominated by horizontal flows that shear and braid field lines over a longer time period. Here, we need to note that the magnetic flux tube in their simulation is initially twisted before it emerges into the solar atmosphere. The observations in \citet{2012Liuyang} show that the magnetic energy in the corona is contributed mainly by vertical flows, while magnetic helicity is mainly contributed by horizontal flows for their investigated ARs.

The AR 12673 on 2017 September 6 produced the largest solar flare (X9.3) in the past decade and was associated with a major coronal mass ejection (CME). About three hours before the flare, another weaker X-class flare took place in the same region. Moreover, according to the investigations of \citet{2017YangSH}, this AR had already produced 12 M-class flares during the two days before the two X-class flares. The occurrences of solar eruptions are related to magnetic energy release. Eruptions associated with X-class flares are generally accompanied by much more magnetic free energy release than other eruptions (e.g., \citealt{2010Jing,2014SuJT}). It is interesting that the occurrences of the two X-class flare eruptions are within only three hours, and the second flare is decade-class as well. Figure \ref{f1}a shows the AR magnetic free energy calculated by subtracting the potential field energy \citep{1978seehafer} from the nonlinear force-free field (NLFFF) energy (see below). The time resolution of the free energy time series is 12 minutes for 18 hours. The energy has a stepwise decrease after the first X-class flare eruption then increases before the major eruption, which implies that the nonpotential energy is accumulated rapidly. This should be related to the continuous flux emergence. The AR experienced the fastest flux emergence three days before the major eruption (see Figure \ref{f1}a). An extraordinary flux emergence rate of $1.12^{+0.15}_{-0.05}\times10^{21} Mx~hr^{-1}$ on September 3 was measured by \citet{2017SunXD}, which accumulates free energy rapidly for the later eruptions. Persistent flux emergence, apparent coalescence, cancelation, and shearing motions are followed. The whole AR is in a rapid evolving state. Total unsigned vertical current is increasing with the unsigned magnetic flux (Figure \ref{f1}b). Helical magnetic structures (i.e., a flux rope) is thought to exist prior to the major eruption \citep{2017YangSH,2018YanXL}. The helicity or twist of the AR can be estimated by the proxy $mean~twist~parameter$ $\alpha$ derived as $\alpha\varpropto\frac{\sum\emph{\textbf{J}}_z\cdot\emph{\textbf{B}}_z}{\sum{\emph{\textbf{B}}_z}^2}$ \citep{1994Pevtsov,2004Hagino}, that has been a popular quantity to correlate with the occurrence of solar energetic events \citep{2014Vemareddy,2016Bobra}. Similar as the unsigned flux, $\alpha$ in Figure \ref{f1}c also exhibits an increasing trend. The increases of the vertical current and the twist parameter $\alpha$ are apparently related to the fast flux emergence.

In this paper, we investigate the buildup of the nonpotentiality in the core of the AR 12673. The evolution of the AR magnetic field by photospheric motions and flux emergence before the major eruption may help us to understand how the magnetic nonpotentiality builds up and what is the dominant contributor to the major eruption. That is important for predicting disastrous space-weather events. The evolution of the eruptive filaments is presented in Section 2 by the EUV observations and the magnetic field extrapolation results. In Section 3 we investigate the nonpotentiality of the AR by analysis of the photospheric magnetic fields. Discussions and conclusions are given in Section 4.

\section{OBSERVATIONS AND MAGNETIC FIELD EXTRAPOLATIONS}
NOAA AR 12673 (S09W42) produced two successive X-class flares on 2017 September 6. The first X2.2 flare started at 08:57 UT, peaked at 09:10 UT, ended at 09:17 UT, and was accompanied by a faint CME. The second X9.3 flare started at 11:53 UT, peaked at 12:02 UT, ended at 12:10 UT, and was accompanied by a strong halo CME. For comparison, two successive X-class flares occurred from the AR 11429 on 2012 March 7 in the same solar cycle \citep{2014Rui}, each of which was associated with a fast CME \citep{2013Liuxying,2014Liuxying}.

\begin{figure}

\epsscale{0.8}
\plotone{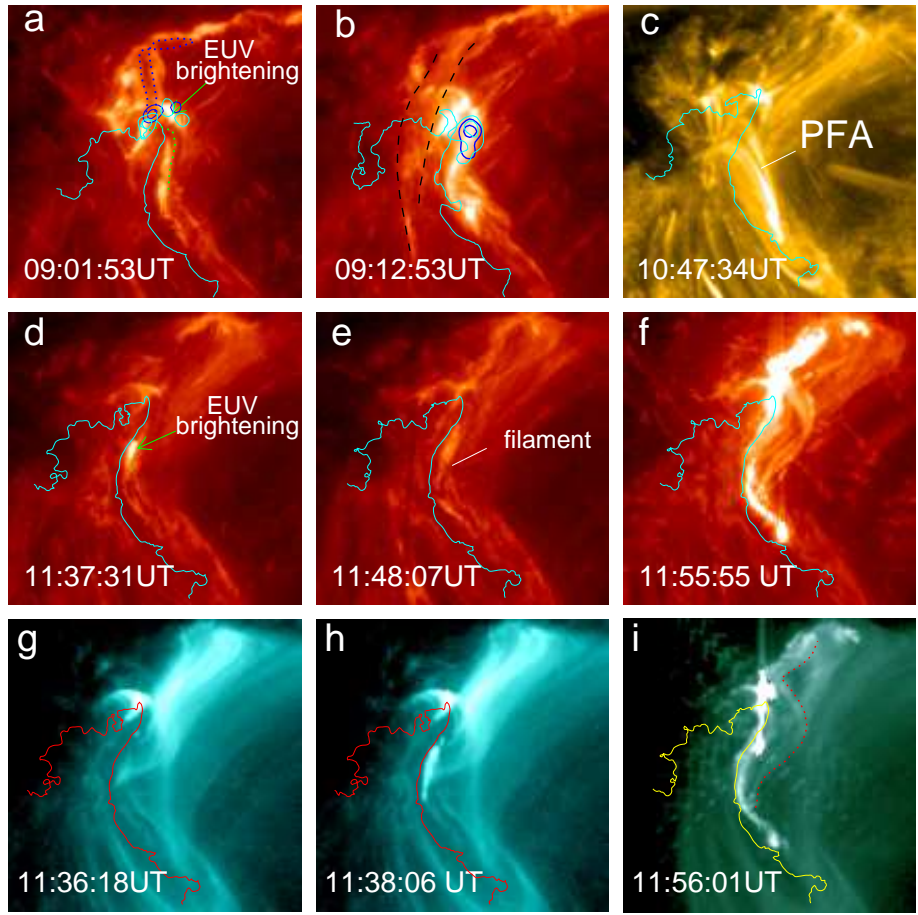}
\caption{EUV observations during the two eruptions. (a) and (b) Evolutions of the filament at AIA 304 \AA~wavelength during the first eruption. The RHESSI sources at 6-12 keV (blue) and  25-50 keV (cyan) are at 09:02 UT in (a) and at 09:10 UT in (b). The blue dotted lines in (a) mark the outline of the first rising structure in the first eruption. The green dotted line in (a) marks the sigmoid structure. The black dashed lines in (b) outline the ejecta. (c) Post flare arcades after the first eruption. (d-f) Evolution of the filament before and during the major eruption. (g) and (h) Overlying arcades before the major eruption at AIA 131 \AA. (i) Overall shape of the rising filament during the major eruption at AIA 94 \AA. The PIL is overplotted by the solid line in each panel.\label{1}}
\end{figure}

Figures \ref{1}a and \ref{1}b exhibit the typical stages of the first eruption using the EUV images in the 304 \AA~channel from the Atmospheric Imaging Assembly (AIA; \citealt{2012Lemen}) on board the Solar Dynamics Observatory (SDO; \citealt{2012Pesnell}). The northern part in Figure \ref{1}a (blue dotted lines) first rises and finally erupts. Figure \ref{1}b shows that the rising structure (black dashed lines) expands gradually. Figure \ref{1}a shows that the EUV brigtenings appear between the northern part (blue dotted lines) and the inverse S-shaped EUV feature (green dotted line). The RHESSI \citep{2002Lin} X-ray sources in Figure \ref{1}a are consistent with the EUV brightenings. It is worth noting that the X-ray sources are spread to the south along the magnetic PIL (RHESSI contours in Figure \ref{1}b) when the flare develops near its peak time. The position of the reconnection signatures changes probably due to involvement of different magnetic field lines into the energy release process (e.g., \citealt{2017Priest,2018Zimovets}). The brightening in Figure \ref{1}a may correspond to the initial reconnection by footpoint motions, and the one in Figure \ref{1}b may be related to the reconnection beneath the rising plasmoid. Unfortunately, RHESSI missed the second major flare. The AIA 171 \AA~in Figure \ref{1}c shows the post flare arcades (PFA) above the PIL. The PFA still looks sheared which is probably inherited from the inner overlying flux tubes of the pre-eruption configuration. \citet{2018YanXL} indicated that a CME emerging from LASCO C2 is associated with the inverse S-shaped structure. Although the AR 12673 experiences a release of magnetic free energy in the first X-class flare, it still possesses strong nonpotentiality.

The middle row in Figure \ref{1} shows the observations at AIA 304 \AA~prior to the major eruption. The EUV brightening (see Figure \ref{1}d) first appears near the PIL. Considering the projection effect (the AR is located at S09W42), we think that the brightening is probably above the PIL. Later on, the filament begins to rise (Figure \ref{1}e). Figure \ref{1}f shows that an EUV structure with helical pattern erupts and the overall region along the PIL between the footpoints of the structure is brightened. The helical structure should be the erupting filament. \citet{2017YangSH} suggested that a kink knot structure exhibited later on near the apex of the helical structure is an observational evidence of the kink instability. Figure \ref{1}g shows an arcade structure at 131 \AA~above the initial brightening. Then, the footpoints of the arcade begins to be brightened (Figure \ref{1}h). Magnetic reconnection probably occurs at these brightening by the interaction between the core fields and the overlying magnetic fields along the PIL. Figure \ref{1}i shows the overall shape of the rising filament at 94 \AA~as shown by the red dotted line. The major eruption seems to occur almost in the same region as the first one after a three-hour time interval, which implies that there exists a large amount of free energy.

\begin{figure}
\epsscale{1}
\plotone{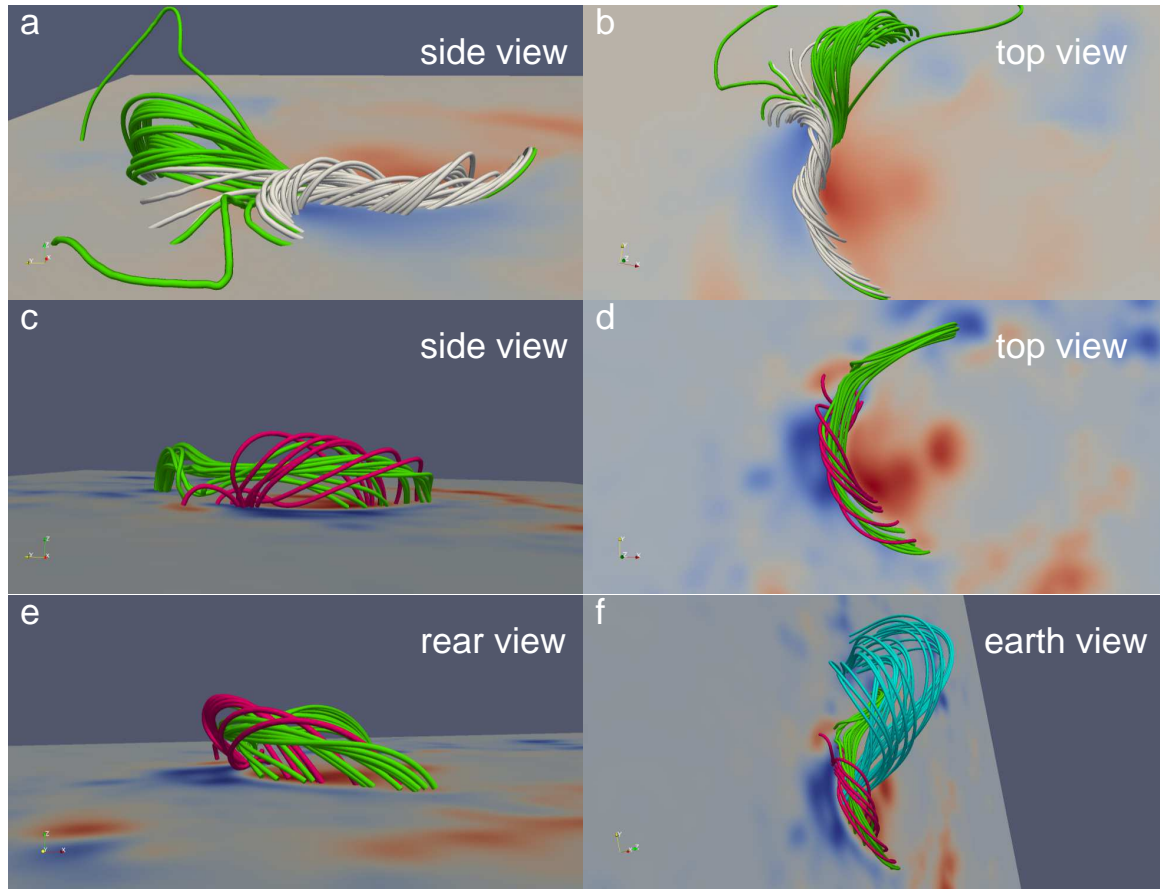}
\caption{Magnetic field extrapolation results of the core fields along the AR PIL. The first row shows the twisted core fields at 08:36 UT from two different views. The middle and bottom rows exhibit the core fields at 11:36 UT from four different views. Overlying fields in cyan are overplotted in (f).\label{2}}
\end{figure}

Figure \ref{2} shows the reconstructed flux-rope structure at different times using the NLFFF extrapolation method \citep{2004Wiegelmann}. Here, we apply this NLFFF method to an area of $280^{\prime\prime}\times220^{\prime\prime}$. The spatial resolution is $1^{\prime\prime}.0$ per pixel. We show some arbitrarily chosen force-free field lines along the PIL to compare with the AIA images at different wavelengths. Figures \ref{2}a and \ref{2}b show the flux-rope structures at 08:36 UT before the first eruption. The flux rope seems to stay lower from a side view. The bundle of grey field lines corresponds to the inverse S-shaped structures (green dotted line) in Figure \ref{1}a. The green field lines match the northern part (blue dotted lines in Figure \ref{1}a) in the EUV image. Figures \ref{2}c-\ref{2}f show the flux-rope structures at 11:36 UT prior to the major eruption. The height of the flux rope seems higher than that before the first eruption, and it is more curved. Figure \ref{2}f shows a bundle of twisted overlying fields (cyan lines) above the flux rope, which seems, in general, to be consistent with the EUV arcades in Figures \ref{1}g and \ref{1}h. The south footpoints of the cyan field lines are just located around the EUV brightening in Figure \ref{1}h. We do not think that the grey sigmoid structure (Figure \ref{2}a) fully erupted during the first eruption, since the twisted rope structure is reconstructed in almost the same region after the eruption. The two eruptions occur successively within only a three-hour time interval. It is possible that the first one is a partial eruption, when just a part of the overall magnetic structure presented has erupted. The observed erupted structures in the EUV images (blue dotted and black dashed lines in Figures \ref{1}a and \ref{1}b) may merely correspond to the bundle of green field lines and parts of the grey lines in Figures \ref{2}a and \ref{2}b. Additionally, the first CME presented by \citet{2018YanXL} was much weaker and slower.

\section{Evolution of Magnetic Nonpotentiality}
\subsection{Distributions of vertical currents and photospheric motions}
\begin{figure}
\epsscale{1}
\plotone{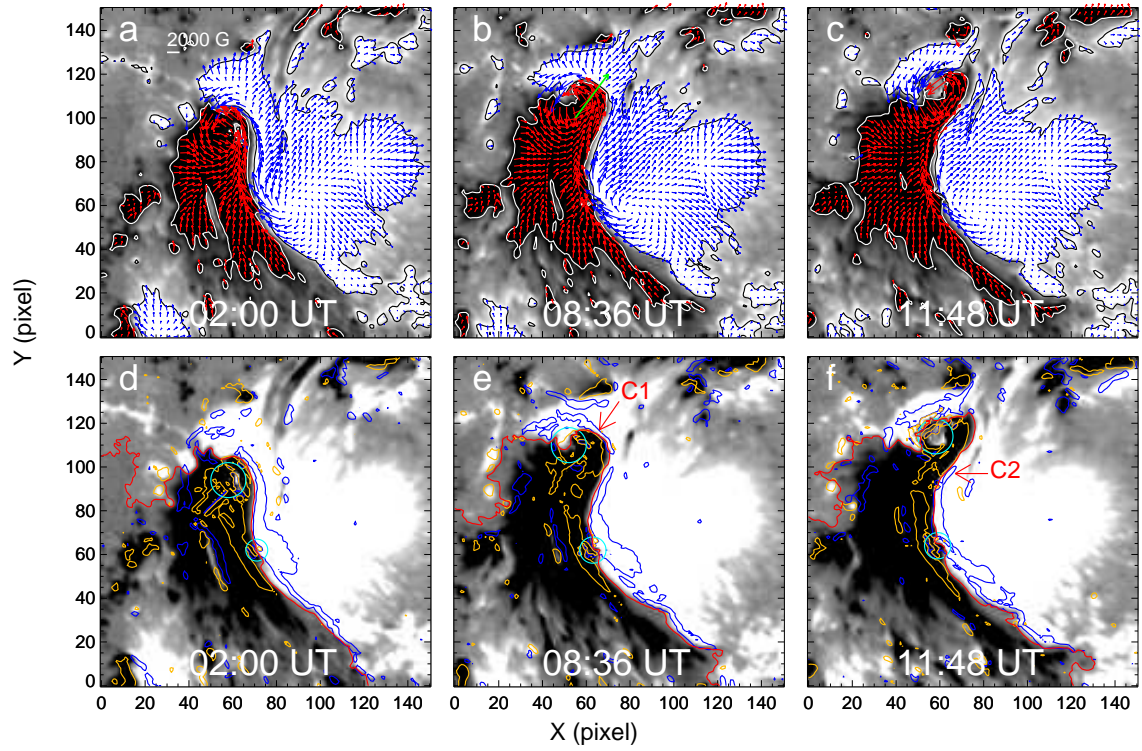}
\caption{Top row: SHARP vector magnetic field in the core region of AR 12673 at specific times. The background image is the normal field with the positive field in white and the negative in black. It is scaled to $\pm 500~G$ (white/black contours). The arrows represent the tangential component of the magnetic field. The colors of the arrows (blue/red) indicate that the normal fields at those pixels are positive/negative. Bottom row: vertical current ($J_z$) distributions overplotted on the corresponding magnetograms. The yellow (blue) contour represents the positive (negative) electric current density $J_z$ of $150$ ($-150$) $mA~m^{-2}$. Cyan circles mark the regions with the pixels of poor quality. The red curve is the PIL.\label{3}}
\end{figure}

The nonpotential nature of magnetic fields is a measure of AR eruptivity and reveals the connection of photospheric magnetic motions with the energy buildup in the AR. The Helioseismic and Magnetic Imager (HMI; \citealp{2012Scherrer,2012Schou}) provides a data product called Space-weather HMI Active Region Patches (SHARPs; \citealp{2014Bobra,2014Hoeksema}). The SHARP data provide vector magnetograms with a pixel scale of about 0$^{\prime\prime}$.5 and a cadence of 12 minutes. Figure \ref{3} shows radial and tangential components of the magnetic fields and vertical currents ($J_z$) distributions in the core field region of the AR 12673. The magnetic transverse vectors are seen to be nearly aligned with the PIL, which implies a large shear along the interface of the two opposite magnetic polarities \citep{2001ZhangHQ}. Owing to the strong shear in the transverse field vectors, the PIL generally spreads with intense vertical current ($J_z$) distribution (e.g., \citealp{2012Vemareddy,2014Sharykin,2014Janvier,2015Sharykin}). The intense $J_z$ here is represented by the yellow and blue contours (vertical currents $J_z^+$ = 150 $mA~m^{-2}$ and $J_z^-$ = -150 $mA~m^{-2}$). From Figures \ref{3}d-\ref{3}f it can be seen that enhanced electric currents are distributed mainly around the PIL (red curve). The intense $J_z$ around the PIL had already existed before 02:00 UT on September 6 (Figure \ref{3}d). High gradients of the vertical current density, due to a change of sign on small spatial scales, appeared a short time before the two eruptions, which implies that the opposite polarities are pressed against each other in a compact configuration (see C1 and C2 in Figures \ref{3}e and \ref{3}f). The field vectors in the top panels are pointing from negative polarity to positive polarity later in the evolution (by the green arrow in Figure \ref{3}b), which indicates significant twist. This gives trouble to the magnetic field disambiguation. Due to the systematic errors and limitations induced by the orbital and other factors, the inversion and disambiguation methods for processing the SHARP data fail in the middle of the umbrae and the north of the negative polarity (pixels of poor quality within the cyan circles).

\begin{figure}

\epsscale{1}
\plotone{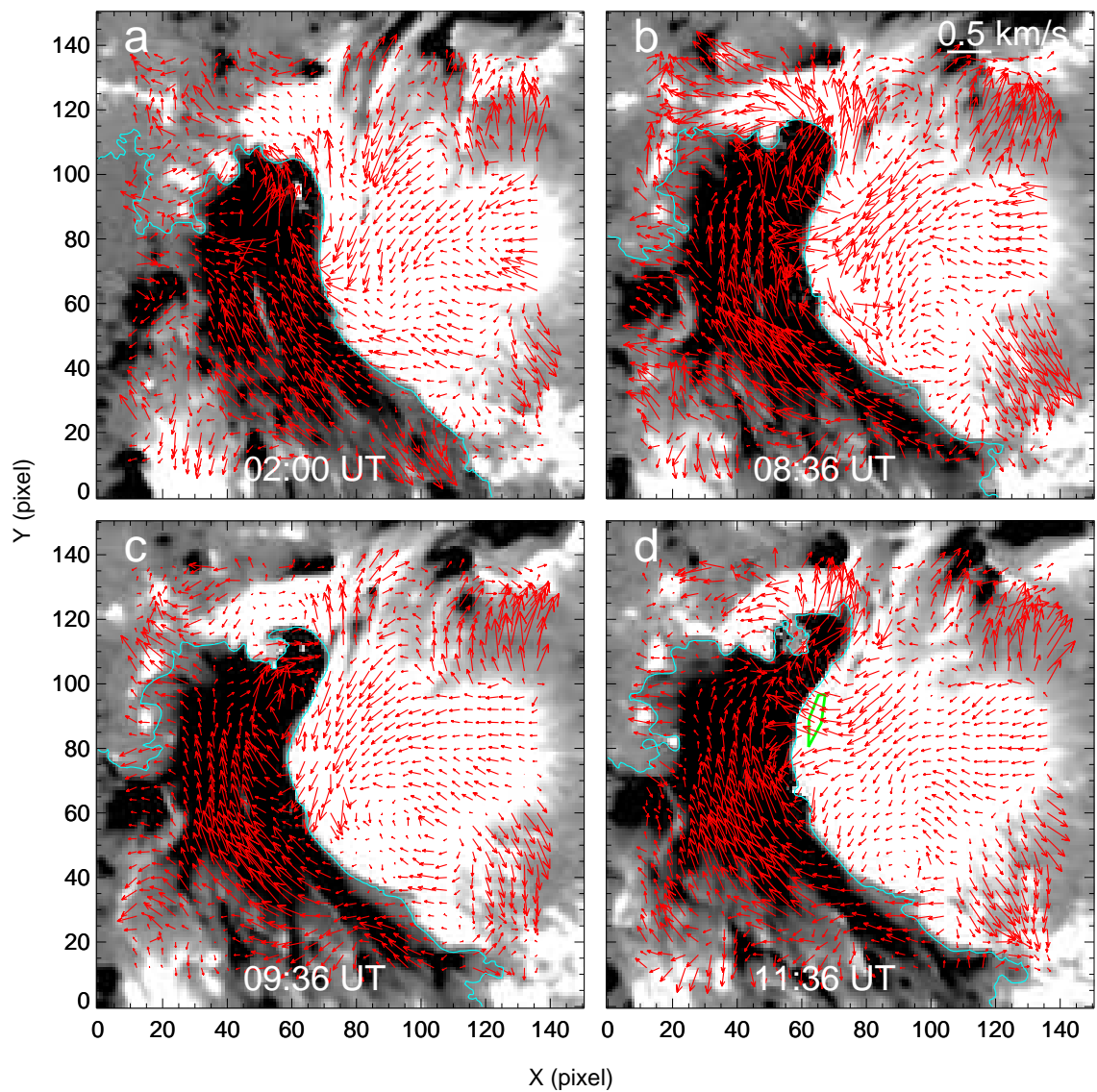}
\caption{Horizontal velocity of flux motions derived from the DAVE4VM technique in the core region of AR 12673 at specific times. The field of view is the same as in Figure \ref{3}. The background image of $B_z$ is scaled to $\pm 500~G$. The green contour in (d) is the outline of the EUV brightening in Figure \ref{1}d. The cyan curve is the PIL.\label{4}}
\end{figure}

Figure \ref{4}a shows that the prominent magnetic motion (derived by DAVE4VM of \citealt{2008schuck}) of the negative polarity is northeastward. It forms shearing motions with the southeastward flows in the positive polarity in the central part of the core region considered. Prior to the first eruption, the motions of the overall region are enhanced as shown in Figure \ref{4}b. The northeastward motion of the negative polarity is getting northward along the PIL. In addition, the southeastward flows in the positive polarity make the two opposite magnetic polarities closer and the PIL more curved. The southward components of the southeastward flows shear with the northward motion of the negative polarity. The northern part of the negative polarity intrudes into the positive one, and forms the so-called ``Magnetic Tongues'' that resemble the $yin~yang$ pattern, which implies the magnetic fields along the PIL are twisted \citep{2011Luoni}. After the first eruption (see Figure \ref{4}c), the southeastward flows in the positive polarity are getting southward, and becomes more parallel to the PIL and more sheared with the northward motion of the negative polarity. About 20 minutes before the major eruption (Figure \ref{4}d), the southeastward flows turn perpendicular to the PIL, and the motion of the negative polarity becomes more parallel to the PIL. It means that the positive polarity is converging to the negative one, and moving perpendicular to the direction of the shearing motion of the negative polarity. This reminds us of the model of \citet{1989Vanballegooijen} to form helical flux-rope structures. Shearing motions produce the sheared magnetic fields, and converging motions make the magnetic shears further increase and cause the reconnection between the sheared field lines. On Figure \ref{4}d we overplot the outline of EUV brightening shown in Figure \ref{1}d. It is just located on the path of the major converging flows near the PIL and near the high gradients of the vertical current density shown in Figure \ref{3}f as C2. Magnetic reconnection may occur around this region. We estimate that the location of the brightening can be above the PIL by assuming the brightening occurs in the chromosphere when considering that the AR is at W42$^\circ$ in longitude.

\begin{figure}
\epsscale{1}
\plotone{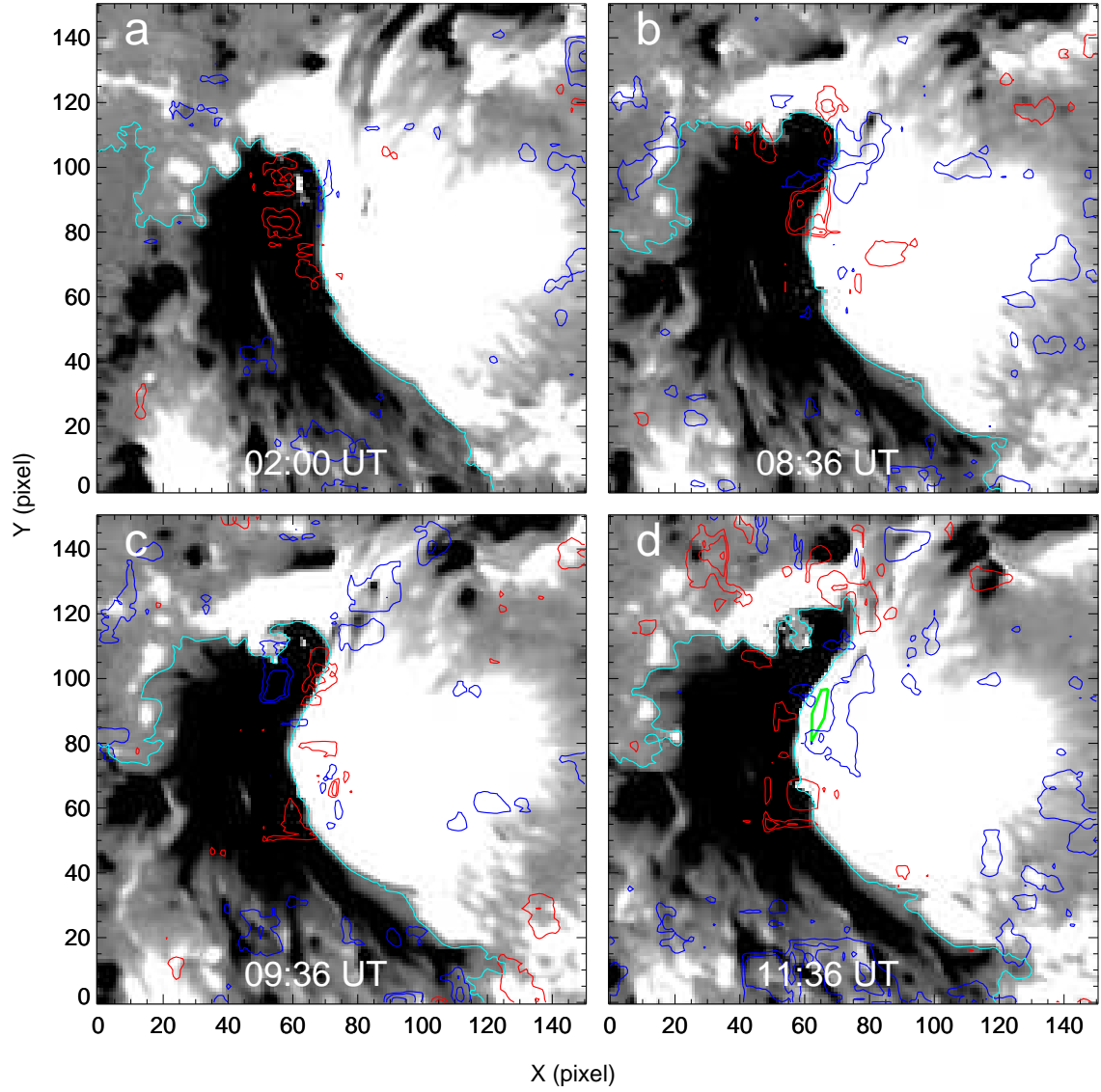}
\caption{Normal velocity of flux motions derived from the DAVE4VM. The blue (red) contours indicate upflows (downflows) of fluxes. The contour levels are $\pm$0.12, $\pm$0.24, $\pm$0.48 $km~s^{-1}$. The background image is the same as Figure \ref{4}.\label{5}}
\end{figure}

We also present the normal velocity derived from DAVE4VM. The blue (red) contours in Figure \ref{5} represent the upflows (downflows). The upflows are related to the flux emergence. Figure \ref{5} shows that the upflows are enhanced before the two eruptions surrounding the major sunspot, which reveals that flux emergence becomes strong before the eruptions. This is consistent with the changes of the total unsigned flux in Figure \ref{f1}a. At 08:36 UT, upflows appear in the northern part of the PIL close to the high gradients of vertical current density C1 in Figure \ref{3}e. Strong downflows also appear along the PIL, which should be related to submergence of magnetic flux, i.e. flux cancelation \citep{1984Rabin,1985Martin}. The submergence of the flux continues even after the first eruption (see Figure \ref{5}c). \citet{1989Vanballegooijen} indicated that the cancelation is probably the result of flux motions. At 11:36 UT (see Figure \ref{5}d), upflows appear next to the projection of the EUV brightening.

It seems that there are some relations between the upflows and the horizontal motions in position. For instance, strong upflows are found at the south of the major northeastward motion of the negative polarity (see Figures \ref{4}a and \ref{5}a). The upflows near C1 is connected to the horizontal flow around C1 (see Figures \ref{4}b and \ref{5}b). Also, the sheared flows along the PIL in the positive polarity at 09:36 UT is connected to the upflows at the north of the sunspot (see Figures \ref{4}c and \ref{5}c). \citet{2012Toriumi} has indicated that the buoyant rise of magnetic field toward the solar surface can drive horizontal flows that are an integral part of the flux emergence process. The magnetic field lines rooted on the photosphere could be dragged with the plasma moving horizontally. Thus, the upflows around the EUV brightening may explain why the continuous southeastward flows in the positive polarity change to perpendicular to the PIL.

\subsection{Magnetic helicity and vertical currents}
Solar eruptions are thought to occur due to an overaccumulation of magnetic helicity (e.g., \citealt{2004Nindos,2012Vemareddy,2014Rui}). Photospheric motions and flux emergence can both inject magnetic helicity by shearing footpoints of magnetic field lines and advecting twisted magnetic flux, respectively. As presented in Figures \ref{4} and \ref{5}, the core region of AR 12673 experiences both relatively strong shearing motions and flux emergence. It is hard to draw an immediate conclusion whether the magnetic helicity is dominated by the shearing motions or emergence of twisted flux. \citet{1984Berger} derived the Poynting-like theorem for the helicity flux through a surface $S$:
\begin{equation}
\frac{dH}{dt}=\int_S2({\bf B}_t\cdot{\bf A}_p)v_ndS+\int_S-2({\bf v}_t\cdot{\bf A}_p)B_ndS \label{dhdt},
\end{equation}
where ${\bf A}_p$ is the vector potential of the magnetic field derived by the SHARP vector magnetogram, which is calculated by means of the fast Fourier transform method as implemented by Chae \citeyearpar{2001Chae}, ${\bf B}_t$ and $B_n$ denote the tangential and normal magnetic fields, $v_n$ and ${\bf v}_t$ are the photospheric normal and tangential velocity fields derived by DAVE4VM. We calculate the two terms in the equation 1, respectively, which is implemented as \citet{2012Liuyang}. The first term indicates that the magnetic helicity in the corona comes from the twisted magnetic flux tubes emerging from the solar interior into the corona (emergence-term hereafter). The second term represents the helicity generated by shearing the field lines by the photospheric motions on the solar surface (shear-term hereafter).

\begin{figure}
\epsscale{1}
\plotone{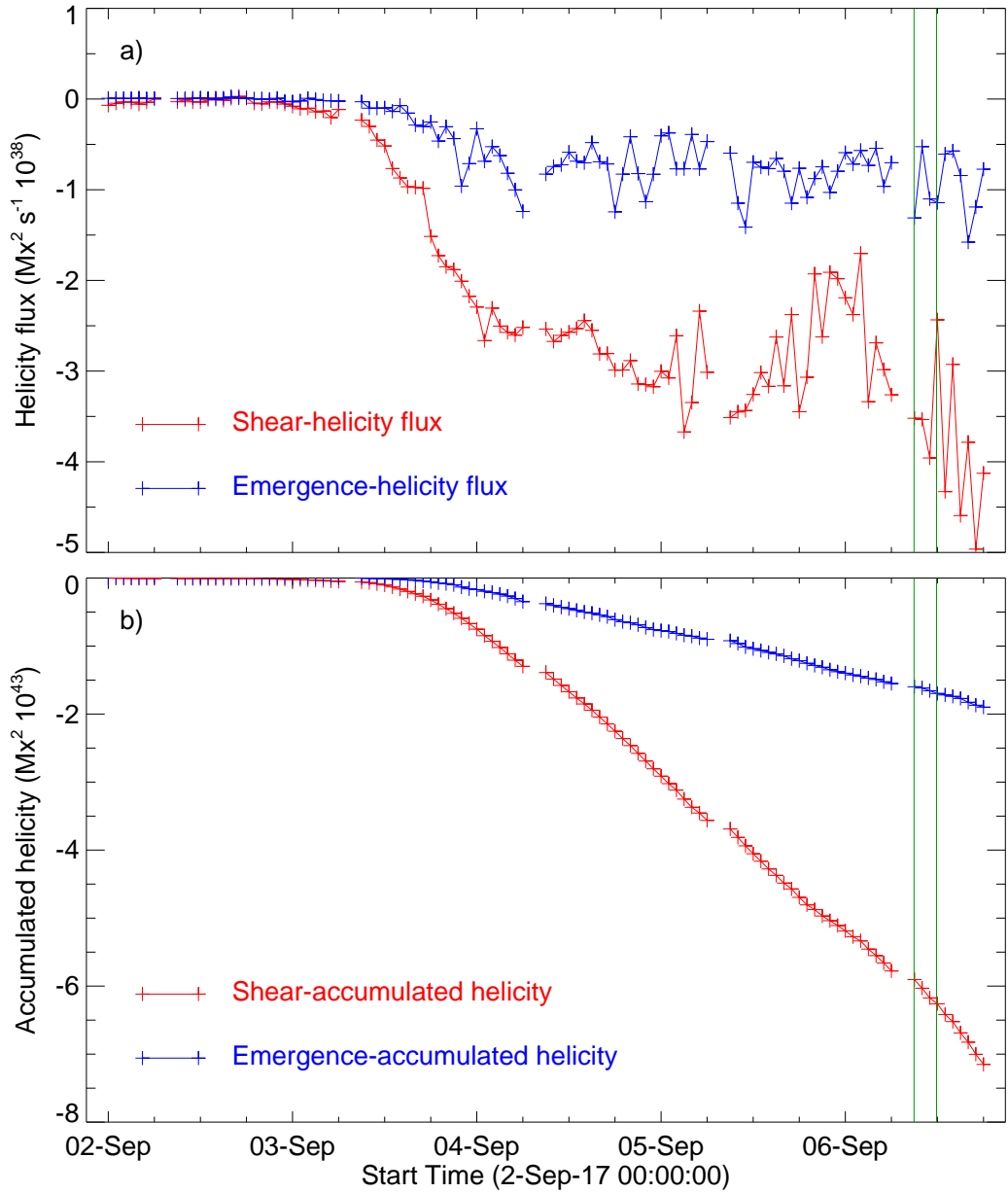}
\caption{Temporal profiles of magnetic helicity of AR 12673. Top panel: red and blue curves represent helicity fluxes across the photosphere from shear and emergence terms, respectively. Bottom panel: red and blue curves refer to accumulated helicities in the corona from shear and emergence terms, respectively. Vertical green lines correspond to the onset times of the X-class flares. \label{6}}
\end{figure}

Figure \ref{6} shows temporal profiles of helicity fluxes across the photosphere and accumulated helicities for the entire AR. Red and blue curves in the top panel represent the helicity flux from $\bf{v}_t$ (shear-helicity flux hereafter) and from ${v}_n$ (emergence-helicity flux hereafter), respectively. Red and blue curves in the bottom panel refer to the accumulated helicities from the shear- and emergence- helicity fluxes, respectively. The accumulated helicity is the integral of the helicity flux over time, which is deemed to be the helicity stored in the corona. We adopt the SHARP vector magnetograms as the input data for calculating the velocity by DAVE4VM and the vector potential ${\bf A}_p$, and the time resolution is 1 hour for 4.75 days (114 hours) from 00:00 UT on September 2. Figure \ref{6}a shows that the shear-helicity flux is dominant starting from September 3. The emergence-helicity flux, on the other hand, remained at a relatively low level for this period. Both helicities are negative. This is in opposition to the so-called hemisphere rule, which predicts that ARs in the southern hemisphere have positive helicity \citep{1995Pevtsov,2014Liuyang1}. The shear-helicity accumulated in the corona before the major eruption is $-6\times10^{43}~Mx^2$; the emergence-helicity accumulated in the corona is only $-1.6\times10^{43}~Mx^2$. The helicities are relatively high in the emerging AR (refer to \citealt{2014Liuyang2}). The shear-term contributes about 79\% of the total helicity, while the emergence-term contributes only about 21\%. Our results imply that most of the magnetic helicity of the AR is probably built up by the horizontal motions. The twisted flux emerging from the subphotosphere contributes to the magnetic helicity much less than the horizontal motions.

Vertical currents in the photosphere are one of the most commonly used measures of magnetic nonpotentiality. Twists or shears by photospheric motions can cause the formation of these currents (e.g., \citealp{1991Melrose,2001ZhangHQ,2015Dalmasse,2016Rui1}). Also, large currents can emerge with the rising current-carrying twisted flux from the subphotosphere (e.g., \citealp{1996Leka,2014Torok,2014Cheung}).
As presented in Figures \ref{3} and \ref{4}, the intense vertical currents are distributed along the PIL where the transverse field vectors are sheared, and probably connected with the photospheric motions. However, it is also possible that the intense currents are carried by the emerging flux. The emergence-helicity can be considered as a proxy for quantifying the amount of emerging current-carrying twisted flux. Thereby, we expect to use the spatial distribution of the helicity flux to identify the major contributor to the intense vertical currents.

\begin{figure}
\epsscale{1}
\plotone{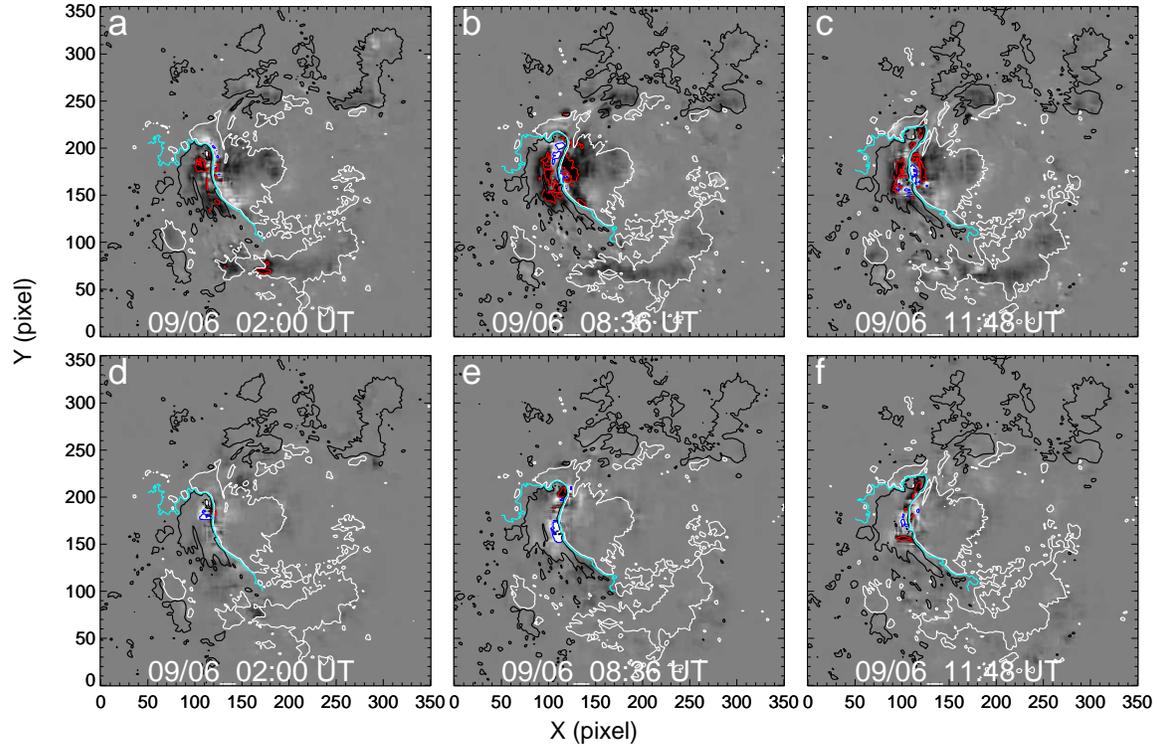}
\caption{Helicity flux distribution ($G_A$) across AR 12673 on specific times. Top and bottom rows correspond to the maps of helicity flux from shear and emergence terms, respectively. Maps are scaled within $\pm10^{20}~Mx^2~cm^{-2}~s^{-1}$. White (black) contours refer to 500 (-500) G levels of $B_n$. Intense signal of positive (negative) helicity flux of $10^{20}$ ($-10^{20}$) $Mx^2~cm^{-2}~s^{-1}$ are represented by blue (red) contours. The cyan curve is the PIL.\label{7}}
\end{figure}

In Figure \ref{7}, we use the spatial distribution of $G_A$ ($dH/dt = \int_S G_A (x,y)dS$, where $G_A = 2[({\bf B}_t\cdot{\bf A}_p)v_n-({\bf v}_t\cdot{\bf A}_p)B_n]$) to represent helicity flux densities before the major eruption. The first and second terms of $G_A$ represent the emergence- and shear-helicity flux densities, respectively. White (black) patches refer to a positive (negative) sign of helicity flux, and we scaled the maps within $\pm10^{20} Mx^2~cm^{-2}~s^{-1}$ so that all magnetic elements are visible. We use the blue (red) contours to represent the intense positive (negative) helicity flux. The top panels show that the intense shear-helicity flux as expected is distributed on both sides of the PIL, and the helicity flux of negative sign is dominant. The intense emergence-helicity flux is also distributed around the PIL, but, in contrast, it is much weaker. Additionally, the emergence-helicity flux calculated includes the flux from both upflows and downflows (refer to the implementations by \citealp{2012Liuyang}). The real helicity flux associated with the emerging current-carrying flux should be less than what we show.

\begin{figure}
\epsscale{1}
\plotone{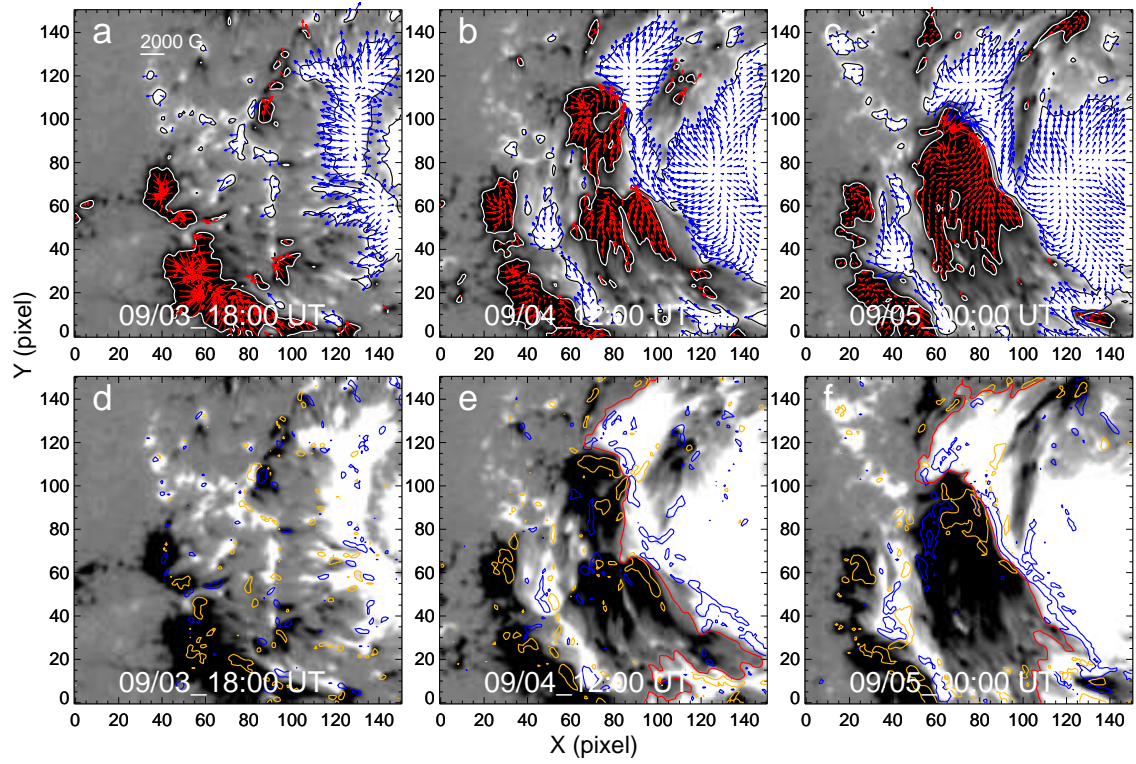}
\caption{Same as in Figure \ref{3}, but for different times.\label{8}}
\end{figure}
\begin{figure}
\epsscale{1}
\plotone{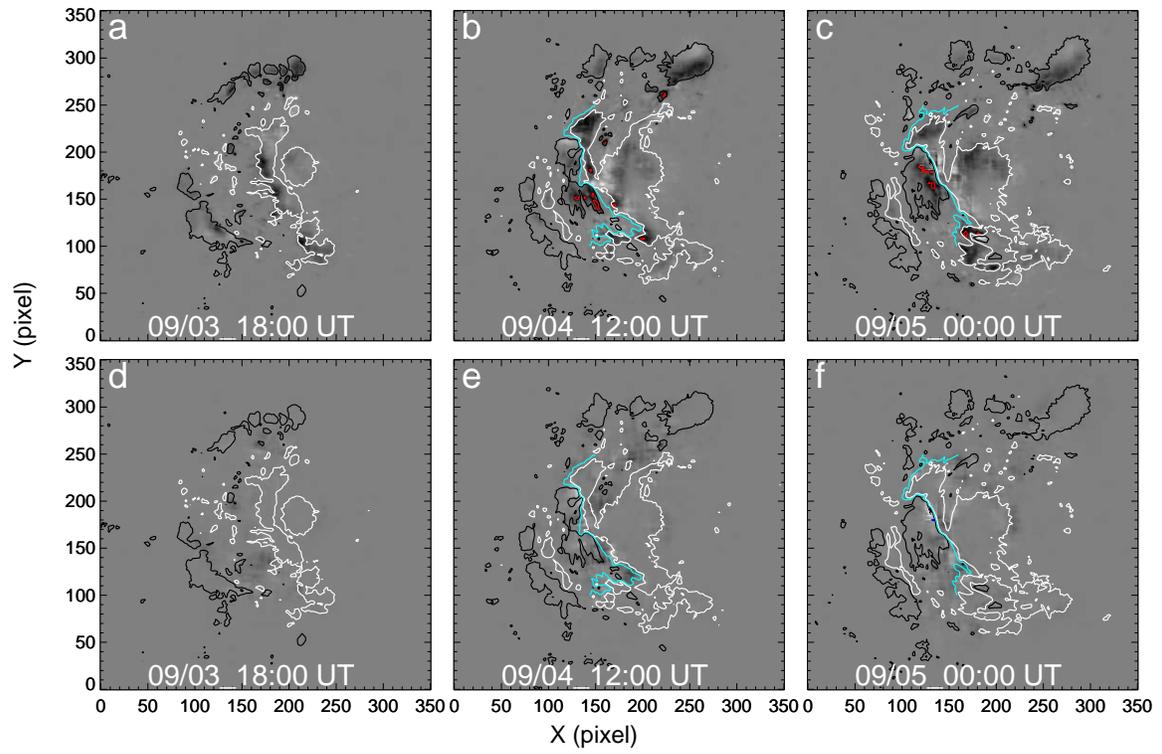}
\caption{Same as in Figure \ref{7}, but for different times (shown in Figure \ref{8}).\label{9}}
\end{figure}

The profile of the shear-helicity flux of the entire AR correlates well with the total unsigned vertical currents (cf. Figures \ref{6} and \ref{f1}b). We make a further investigation on the relation between the shear-helicity and the distribution of the vertical currents, and present the distributions of the vertical currents and vector magnetic fields in the earlier phase of the emergence in Figure \ref{8}. On September 3, the AR experienced the fastest flux emergence, and the distribution of the vertical current was dispersed (see Figures \ref{8}a and \ref{8}d). When two major opposite polarities begin to touch, intense vertical currents appear around the touching places. With more parts of the two polarities are pressed against each other, longer intense current contours appear on both sides of the PIL. Obviously, the intense vertical currents are formed due to the converging motions that press the opposite polarities together in a compact configuration. The magnetic field strength is enhanced and the magnetic fields become sheared along the PIL. On the other hand, we also present the distributions of helicity fluxes at the corresponding times in Figure \ref{9}. In order to compare with the helicity fluxes in Figure \ref{7}, we still scaled the maps within $\pm10^{20} Mx^2~cm^{-2}~s^{-1}$. The shear-helicity flux is still dominant in the earlier phase of the flux emergence. The intense shear-helicity flux distributed around the PIL increases with the content of the intense vertical currents, which also reveals the close relation between the photospheric motions and the vertical currents.

\section{Discussion and conclusion}

AR 12673 produced very strong energy release on 2017 September 6 that has drawn great scientific attention due to the intense flare as well as the significant impact on the space weather. One of the most prominent features is the fast flux emergence process that even continues after the major eruption, and the photospheric magnetic fields experience fast evolving processes. Using HMI SHARP vector magnetic field data, we studied the helicity in the corona. The emergence-helicity accumulated in the corona is $-1.6\times10^{43}~Mx^2$ before the major eruption, while the shear-helicity accumulated in the corona is $-6\times10^{43}~Mx^2$, which contributes about 79\% of the total helicity. The shear-helicity flux is dominant from the beginning of the flux emergence to the major eruption, and the situation continues after the eruption. This is consistent with the conclusion of \citet{2012Liuyang} that the magnetic helicity is mainly contributed by the horizontal flows for their investigated ARs. However, our observational results are different from the results obtained by the MHD simulations \citep{2004Manchester,2006Magara,2007Manchester,2009FanYH,2012Fangfang}, that the emergence-helicity is dominant in the early emergence phase, and the shear-helicity is dominant later. Our results imply that the emerged fields initially contain relatively low helicity. Much more helicity is built up by shearing and converging flows acting on the preexisted or emerging flux. We tend to believe that parts of the shearing flows come from horizontal divergent flow due to the rising flux in the subphotosphere, since the profile of the shear-helicity flux correlate well with the total unsigned flux, and the distributions of horizontal flows of the flux appear to connect with the vertical flows.

Our results show that the intense vertical currents are mainly distributed along the PIL where the magnetic fields are sheared. The evolution of the vertical currents also shows that most of the intense currents do not appear initially with the emergence of the flux, which implies that this emerging flux is probably not strongly current-carrying, i.e., field lines are not twisted strongly in the early emergence phase. This is consistent with our conclusion from the study of the magnetic helicity. On the other hand, we find that the intense vertical currents appear on both sides of the PIL as two opposite magnetic polarities are pressed against each other. Shearing motions are getting stronger with the flux emergence, especially on both sides of the PIL of the core field region formed by the coalescence of the magnetic polarities after September 3, which is reflected in the shear-helicity flux. Therefore, the vertical currents on both sides of the PIL in the core field region are mainly contributed by the photospheric motions. The presence of intense vertical current quantifies the twist and shear of the magnetic fields. Accordingly, we think that the flux rope in the core field region, exhibited by the extrapolations, is formed by long-term photospheric motions. The converging motions that press the opposite polarities against each other enhance the extent of magnetic shears and the gradient of magnetic fields. \citet{2001ZhangHQ} found that the shear and gradient of the magnetic field are important and reflect a part of the electric current in solar ARs.

\begin{figure}
\epsscale{0.8}
\plotone{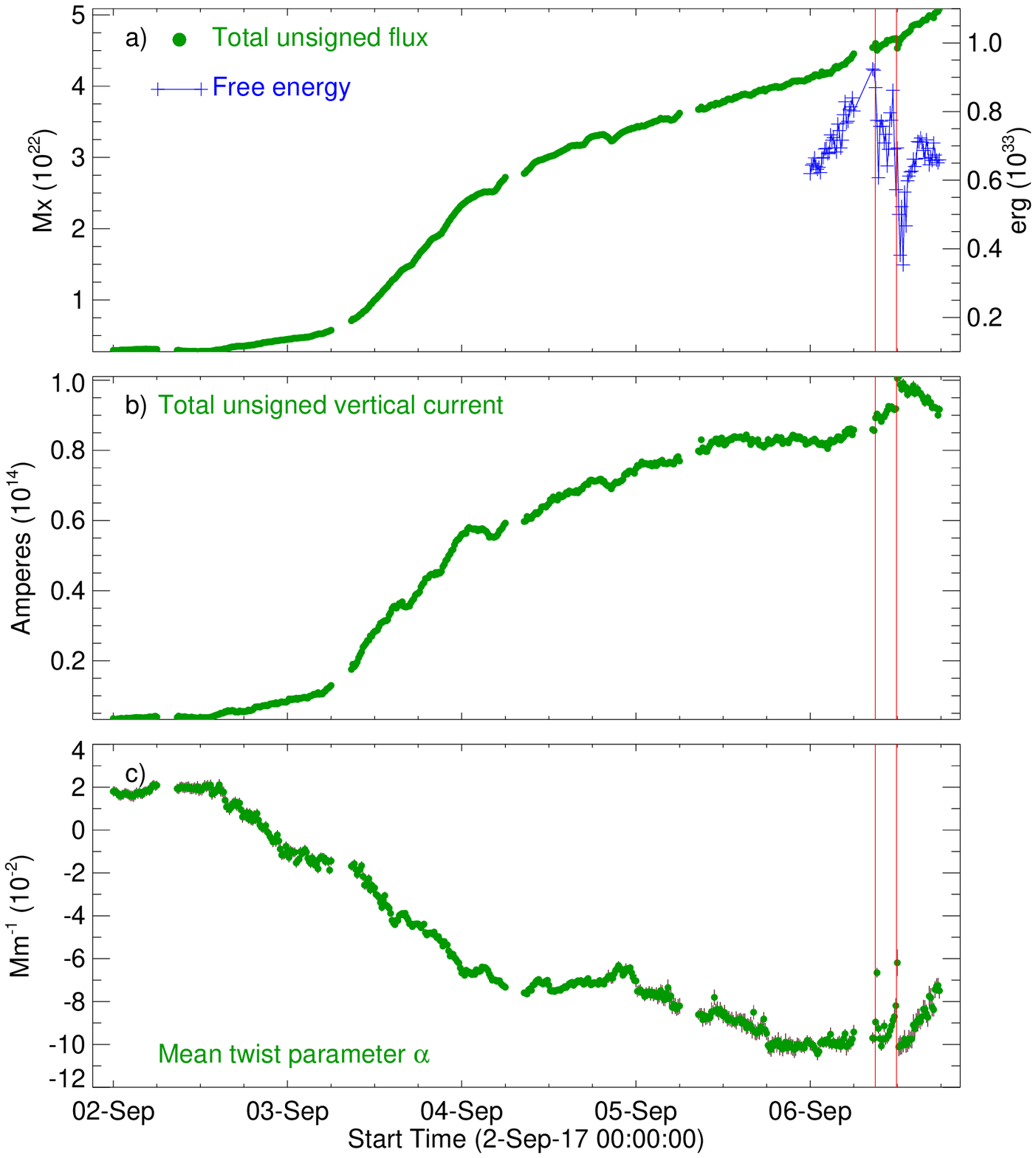}
\caption{Time evolution of magnetic parameters in the AR 12673. These parameters available on a twelve-minute cadence are provided by the keywords in the SHARP data series, and used for the AR event forecasting and for identifying regions of interest \citep{2014Bobra,2016Bobra}. The pixels that contribute to these parameters are selected by examining two data segment maps: BITMAP ($\ge$ 30) and CONF\_DISAMBIG (disambiguation noise threshold $\approx$ 150 G, CONF\_DISAMBIG = 90, \citealt{2014Bobra}) in the SHARP data. (a) Total unsigned flux and magnetic free energy by magnetic field extrapolations. (b) Total unsigned vertical current. (c) Mean twist parameter $\alpha$. The vertical red lines represent the onset times of the X-class flares. Uncertainties are overplotted for each of the parameters. \label{f1}}
\end{figure}

The shearing and converging motions are continuously generated driven by the flux emergence. Magnetic reconnection probably occurs in the way of \citet{1989Vanballegooijen} by shearing and converging motions. The twist of the magnetic field increases with time. Figure \ref{f1}c shows that the mean twist $\alpha$ of the entire AR reach a high value of $\sim-0.10~Mm^{-1}$ before the eruptions. For comparison purposes, we also calculated $\alpha$ of other ARs with intense eruptions, e.g., AR 11429: $\sim-0.05~Mm^{-1}$, AR 11158: $\sim0.03~Mm^{-1}$, AR 12192: $\sim-0.01~Mm^{-1}$. Although the AR has a strong twist before the eruption, this cannot indicate that the eruption is triggered by the kink instability, since $\alpha$ has maintained around the high value for a dozen hours before the eruptions. Therefore, some other mechanisms are necessary to be considered. Photospheric motions or flux emergence is the potential trigger for the major flare and eruption.

The AR 12673 is a representative as the photospheric motions contribute most of the nonpotentiality in the AR with vigorous flux emergence. The knowledge to the role of photospheric motions and flux emergence is important for predicting disastrous space-weather events.

\acknowledgments
The work was supported by NSFC under grants 41604146, 41774179, and 41374173, the Specialized Research Fund for State Key Laboratories of China, and Strategic Priority Program on Space Science (No. XDA 15011300). The authors gratefully acknowledge financial support from the study abroad program of the Chinese Academy of Sciences. The data used here are courtesy of the NASA/SDO HMI and AIA science teams. We thank the HMI science team for making the SHARPs vector magnetograms available to the solar community.

%



\clearpage

\end{document}